\documentclass[twocolumn]{aastex701}
\NewPageAfterKeywords
\usepackage{amsmath}
\usepackage{multirow}

\newcommand{\teff}{$T_{\text{eff}}$}
\newcommand{\trb}{$T_{\text{RB}}$}

\newcommand{\vsini}{$v\sin{(i)}$}

\newcommand{\rprs}{$R_{\text{p}}/R_{\star}$}

\newcommand{\kms}{$\text{km }\text{s}^{-1}$}
\newcommand{\ms}{$\text{m }\text{s}^{-1}$}

\shorttitle{Obliquities of Four S-Type Hot Jupiters}
\shortauthors{Polanski et al. 2026}
\graphicspath{{./}{figures/}}
\begin{document}

\title{CHIANTE I: Obliquity Measurements of Four High-Priority Ariel Targets in Binaries}

\correspondingauthor{Alex S.\ Polanski}
\email{aspolanski@lowell.edu}

\author[0000-0001-7047-8681]{Alex S. Polanski} 
\altaffiliation{Percival Lowell Fellow}
\affil{Lowell Observatory, 1400 W Mars Hill Road, Flagstaff, AZ, 86001, USA}
\email{aspolanski@lowell.edu}

\author[0000-0002-7670-670X]{Malena Rice}
\affiliation{Department of Astronomy, Yale University, 219 Prospect Street, New Haven, CT 06511, USA}
\email{malena.rice@yale.edu}

\author[0000-0002-2361-5812]{Catherine A. Clark}
\affiliation{NASA Exoplanet Science Institute, IPAC, MS 100-22, Caltech, 1200 E. California Blvd., Pasadena, CA 91125, USA}
\email{clarkc@ipac.caltech.edu}

\author[0000-0002-9288-3482]{Rachael M.\ Roettenbacher}
\affiliation{Department of Astronomy, University of Michigan, 1085 S.\ University Ave., Ann Arbor, MI 48109, USA} 
\email{rmroett@umich.edu}

\author[0000-0002-3852-3590]{Lily L.\ Zhao}
\affiliation{Flatiron Institute, Simons Foundation, 162 Fifth Avenue, New York, NY 10010, USA}
\email{lilylingzhao@uchicago.edu}

\author[0000-0002-9873-1471]{John M.\ Brewer}
\affiliation{Dept. of Physics \& Astronomy, San Francisco State University, 1600 Holloway Ave., San Francisco, CA 94132, USA}
\email{jmbrewer@sfsu.edu}

\author[0009-0007-0922-7315]{Momo Ellwarth}
\affiliation{Lowell Observatory, 1400 W Mars Hill Road, Flagstaff, AZ, 86001, USA}
\affiliation{Department of Astronomy and Planetary Science, Northern Arizona University, PO Box 6010, Flagstaff, AZ 86011 USA}
\email{mellwarth@lowell.edu}

\author[0000-0002-0388-8004]{Emily A.~Gilbert}
\affiliation{NASA Exoplanet Science Institute, IPAC, MS 100-22, Caltech, 1200 E. California Blvd., Pasadena, CA 91125, USA}
\email{emily.a.gilbert@jpl.nasa.gov}

\author[0000-0000-1234-5678]{Joe Llama}
\affiliation{McDonald Observatory, University of Texas at Austin. 2515 Speedway. Austin, TX 78712, USA}
\email{joe.llama@utexas.edu}

\author[0000-0002-7216-2135
]{Andrew W. Mayo}
\affiliation{Dept. of Physics \& Astronomy, San Francisco State University, 1600 Holloway Ave., San Francisco, CA 94132, USA}
\email{mayo@sfsu.edu}

\author[0000-0002-4974-687X]{Andrew E.\ Szymkowiak}
\affiliation{Department of Astronomy, Yale University, 52 Hillhouse Ave., New Haven, CT 06511, USA}
\affiliation{Department of Physics, Yale University, 217 Prospect St, New Haven, CT 06511, USA}
\email{andrew.szymkowiak@yale.edu}

\author[0000-0002-8552-158X]{Gerard T. van Belle} 
\affil{Las Cumbres Observatory, 6740 Cortona Drive, Suite 102
Goleta, CA  93117, USA}
\affil{Department of Astronomy and Planetary Science, Northern Arizona University, PO Box 6010, Flagstaff, AZ 86011 USA}
\email{gerard.van-belle@nau.edu}

\author[0000-0002-5870-8488]{Olivia Weiss}
\affil{Lowell Observatory, 1400 W Mars Hill Road, Flagstaff, AZ, 86001, USA}
\email{oweiss@lowell.edu}

\submitjournal{AJ}

\begin{abstract}

We present the first results from CHIANTE: a program using the EXtreme PREcision Spectrograph (EXPRES) at the Lowell Discovery Telescope to characterize potential targets of the \textit{Ariel} mission, anticipated to launch in 2031. We report Rossiter-McLaughlin measurements of four Ariel tier 3 hot-Jupiters which reside in binary star systems: KELT-2 Ab, KELT-3 Ab, TOI-1333 Ab, and TOI-1789 Ab. Joint modeling of EXPRES and archival radial velocities with photometry from TESS finds all four planets to be aligned their host stars, despite the host stars spanning the \teff~realignment break, which has been found to divide the planets in multi-star systems into two subsets: those around cool stars that are preferentially aligned, and those around hot stars that exhibit stellar obliquities consistent with isotropy. We revise the \teff~realignment break to be $=6193\pm103$ K, consistent with, but hotter than, previous work. We compare the observed stellar obliquity distribution for all multi-star, hot-Jupiter hosts above this boundary to an expected distribution produced via stellar von-Zeipel-Kozai-Lidov (ZKL) oscillations, a mechanism often invoked to explain misaligned planets in multi-star systems. A simple population synthesis model finds that a pure ZKL population is unable to replicate the observed obliquities. In particular, both the number of aligned and near-polar systems we see today are underestimated. However, adding contributions from aligned and planet-planet scattering populations alongside ZKL oscillations better describes the observed distribution. Nonetheless, more obliquity measurements for planets in multi-star systems are needed to better discern the contributions of each mechanism the observed stellar obliquity distribution.

\end{abstract}

%% Keywords should appear after the \end{abstract} command. 89+


\section{Introduction} \label{sec:intro}

Since the discovery of the first planets outside our Solar System \citep{WolszczanFrail1992,MayorQueloz1995}, a steady rate of discovery has resulted in more than 6,000 confirmed worlds \citep{Christiansen2025}. The success of JWST and its ability to probe the atmospheres of many of these exoplanets means we are now firmly in an era of planet characterization, rather than just discovery. Observations of gas giants have revealed signatures of CO, CO$_{\text{2}}$, CH$_{\text{4}}$, and H$_{\text{2}}$O, which are shedding light on the structure and dynamics of these atmospheres \citep[e.g.][]{Feinstein2023,Carter2024,Carone2023,Inglis2024}. Unexpected surprises, like the detection of sulfur compounds (SO$_{\text{2}}$, H$_{\text{2}}$S), are providing new ways to connect volatile content to planet formation beyond the carbon-to-oxygen ratio \citep{Tsai2023,Powell2024,Crossfield2023,Fu2024,Zhang2025}. To date, JWST has, or is expected to, observe $\sim$80 hot- and warm-Jupiters \citep[according to the TrExoList accessed July 19, 2026;][]{trexolist}. While this number will certainly increase over the mission's lifetime, a dedicated platform for studying exoplanet atmospheres is necessary to survey more than just a fraction of short period giants and explore the properties of these gaseous worlds at population scale. The European Space Agency's fourth medium-class mission, \textit{Ariel}, will fill this niche by surveying 1,000 planet atmospheres \citep{Tinetti2018}, the majority of which will be gas giants.

\textit{Ariel}'s stratification of targets into three tiers is intended to enable not only survey-level studies but also the detailed study of individual planets \citep{Zingales2018,Edwards2019}. The Tier 1 reconnaissance survey of $\sim$1,000 planets will address questions about the exoplanet population as a whole, such as the fraction of planets that have cloudy atmospheres, or those that still retain primordial envelopes of hydrogen/helium \citep{Tinetti2022}. These observations will eventually yield the Tier 3 targets that represent the best opportunities to conduct detailed studies of individual planets' atmospheric compositions, temporal variability, and dynamics. Leading up to launch, \textit{Ariel}'s mission reference sample (MRS) will be optimized in order to maximize the scientific return of the mission. A critical input to this process will be a large mission candidate sample \citep[MCS,][]{Zingales2018} that offers a diversity of targets for which \textit{Ariel} can achieve a high signal-to-noise ratio (SNR), while also addressing multiple science cases. Many known planets satisfy these requirements, but ongoing transit surveys -- in particular the Transiting Exoplanet Survey Satellite \citep[TESS,][]{Ricker2015} -- continue to provide thousands of candidate exoplanets around bright stars which may yet prove fruitful for \textit{Ariel}'s prime survey \citep{Edwards2019}. 

Since the \textit{Kepler}/\textit{K2} mission \citep{Borucki2010} -- and throughout the TESS era -- a strong, global community has grown in support of the transit surveys, bringing numerous ground-based facilities to bear in an effort to confirm exoplanet candidates \citep{Covino2013,Reiners2018,TFOP2018,MTS1,Howell2021,Clark2022,TKS0,Polanski2024}. NASA's participation in \textit{Ariel} through the Contribution to \textit{Ariel} Spectroscopy of Exoplanets \citep[CASE,][]{Zellem2019} module will not only provide optical photometry, but will also introduce avenues for the U.S. follow-up community to participate in the success of the mission. The U.S. Contributions to \textit{Ariel} Preparatory Science (US-CAPS) program is one such conduit that enables collection of the data needed to confirm new TESS Objects of Interest (TOIs) and refine the parameters of known planets and their host stars.% This is especially relevant for targets only accessible in the northern hemisphere, where the majority of these observing resources are concentrated.

This is the first in a series of papers presenting observations from the EXtreme PREcision Spectrograph \citep[EXPRES,][]{Jurgenson2016,Blackman2020} in support of the follow-up effort of \textit{Ariel} targets. Here, we report obliquity measurements of four \textit{Ariel} target hot-Jupiters that orbit one of two stars in a binary-star system (circumstellar or S-type planets). In \S \ref{sec:chiante} we describe the scope and goals of our observational program. \S \ref{sec:observations} describes the observations collected using EXPRES and Lowell Observatory's 1-meter telescope, as well as the archival data used in the joint modeling is described in \S \ref{sec:analysis}. In \S \ref{sec:discussion}, we place these systems in the overall obliquity distribution of planets in multi-star systems and examine various formation and migration pathways that best describe this distribution.

\section{The CHIANTE Program}\label{sec:chiante}

The CHaracterization of prIority Ariel Northern Targets with EXPRES (CHIANTE) program is an effort to confirm and characterize potential \textit{Ariel} targets using the EXtreme PREcision Spectrograph (EXPRES). Fed by the 4.3-meter Lowell Discovery Telescope \citep[LDT,][]{Levine2012}, EXPRES is a an \'echelle spectrograph with  a median resolution of $\sim$137,000 across a wavelength range of 380-780 nm \citep{Jurgenson2016}. It it capable of 10 cm s$^{-1}$ internal instrumental precision and has demonstrated stability of 50-80 cm s$^{-1}$ over long timescales on quiet stars\citep{Brewer2020,Blackman2020,ESSPII}. In support of the EXPRES observations, we are employing additional resources at Lowell Observatory, including speckle imaging from the Quad-Wavefront sensing Six-Channel Speckle Interferometer \citep[QWSSI,][]{QWSSI} and timeseries photometry from the 1-meter Peggy and Eric Johnson telescope \citep[PJ1M,][]{Hardesty2024}. CHIANTE aims to provide data products to address a number of science cases, detailed in the following sections. 

\subsection{Stellar Characterization}

The physical properties of planets are derived from stellar mass, radius, and effective temperature, and therefore the precision of planetary properties depends on our ability to precisely determine stellar parameters. Multi-band photometry can provide rough estimates of these parameters but can leave additional important stellar properties (e.g., metallicity) unconstrained. High-resolution spectroscopy, in combination with parallaxes from \textit{Gaia} \citep{gaia} and isochronal analysis, has become the gold standard for stellar characterization, routinely providing $R_{\star}$, $M_{\star}$, and \teff{} to within a few percent \citep{Berger2023}. Additionally, high-resolution spectra allow for the measurement of precise elemental abundances, providing the appropriate stellar context for the atmospheric composition measurements carried out by JWST, ground-based instruments, and soon \textit{Ariel} \citep{Adibekyan2012,Brewer2016a,Rice2020,Polanski2022,Kolecki2021,Pelletier2025,Freckleton2025}. Furthermore, $\alpha$-element abundances relative to iron provide information in addition to \textit{Gaia} velocities in enabling the study of the exoplanet population across different galactic populations \citep{Santos2017,Clark2021,Behmard2025}.

CHIANTE will obtain high-resolution stellar spectra with EXPRES for a subset of TOIs and known planet-hosts to provide precise stellar parameters and elemental abundances. This catalog will be presented in a subsequent publication.
    
\subsection{Candidate Confirmation and Masses} Independent detection of transiting planet candidates via radial velocity measurements is typically the final step in the planet confirmation process. However, radial velocity follow-up can be time-intensive, leading to a bias in confirmation efforts toward TOIs deemed most scientifically promising on an individual basis. Although the \textit{Ariel} MCS includes a diversity of planet types, it is predominantly composed of hot- and warm-Jupiters \citep{Zingales2018,EdwardsTinetti2022}. These planets typically exhibit larger scale heights, making them ideal for a survey mission such as \textit{Ariel}, but less likely to be the primary result of any individual confirmation study. Large catalogs of these planets are relieving the bottleneck \citep{Schulte2024,Yee2025}, but many more candidates await confirmation.

In addition to confirmation, radial velocities also provide constraints on planet mass. A precise mass measurement at 5$\sigma$-precision or greater is seen as a necessary prerequisite to atmospheric characterization to keep an uncertain mass from dominating the error budget \citep{Batalha2019}. Inferring a mass through the height of spectral features may allow for a less precise mass measurement for gas giants, although this depends heavily on the assumed nature of the atmosphere \citep{DiMaio2023}. Furthermore, the mass-radius relation plateaus near one Jupiter radius for an order of magnitude in mass, making it difficult to prioritize the best targets based on radius alone.

Through US-CAPS, EXPRES will contribute radial velocity measurements for approximately two dozen TOIs, the majority of which are close-in gas giants. The masses provided by CHIANTE will help identify the best planets for consideration in the MRS.

\subsection{Stellar Obliquities} 

Connecting atmospheric composition back to planet formation theory is notoriously fraught with many uncertain factors, including the ambiguity of how the planets we see today migrated to their current locations. For hot Jupiters, stellar obliquities can identify specific migration paths (e.g., through a disk or post-disk-dispersal avenue), making this quantity a valuable parameter against which to compare atmospheric composition. BOWIE-ALIGN \citep{Penzlin2024,Kirk2024} is conducting a survey of a small sample of hot Jupiters with JWST. \textit{Ariel} will expand this sample enormously, with the bottleneck now becoming the number of obliquity measurements.

CHIANTE will provide obliquity measurements for the planets most likely to be targeted by \textit{Ariel} through Rossiter-McLaughlin observations \citep[RM,][]{Rossiter1924, McLaughlin1924}. While not the primary deliverable, CHIANTE will join other surveys that are providing obliquities across a wide range of parameter space in order to trace the formation and migrational history of a larger sample of planets \citep[e.g. ][]{posideonI,kpf_slopes,atreidesI}.

\section{Observations \& Archival Data}\label{sec:observations}

\subsection{EXPRES Spectroscopy}

In-transit spectra were obtained using EXPRES at the LDT near Flagstaff, AZ. Exposure times for each target were determined by balancing the appropriate SNR required for precision RVs, while also achieving good sampling of the transit. Each observing run was timed to allow for approximately an hour of pre- and/or post-transit baseline. Spectra of a Thorium-Argon (ThAr) lamp and laser frequency comb were taken every 30 minutes for wavelength calibration. A summary of the EXPRES observations is given in Table \ref{tab:obs_stats}, and a sample of the EXPRES radial velocities are given in Table \ref{tab:rv_sample}.

\begin{deluxetable*}{lcccr}[t!]
\label{tab:obs_stats}
\tablecaption{Summary of EXPRES  Observations}
\tablehead{\colhead{Target} &
\colhead{Date (UTC)} & \colhead{No. Spectra} & \colhead{Exposure Time [s]} & \colhead{Notes}
}
\startdata
KELT-2 A & 2025-11-30 & 40 & 600 & Seeing 1\farcs6-1\farcs8. Light cirrus.\\
\hline
KELT-3 A & 2025-01-19 & 23 & 800 & Seeing 2\farcs0. Intermittent clouds. \\
KELT-3 A & 2025-02-26 & 21 & 800 & 0\farcs9-1\farcs0. Clear. \\
\hline
TOI-1333 A & 2025-08-08 & 24 & 900 & 1\farcs3-1\farcs5. Clear. \\
\hline
TOI-1789 A & 2026-03-04 & 21 & 850 & 1\farcs3. Clear. \\
TOI-1789 A & 2026-03-19 & 18 & 850 & 1\farcs0. Clear. \\
\enddata
\end{deluxetable*}

Raw spectra were reduced and radial velocities calculated using the cross-correlation function (CCF) method according to \cite{Petersburg2020}. An empirical telluric model was generated using SELENITE \citep{Leet2019}. We also corrected the reduced spectra for chromatic changes caused by variations in airmass with a prescription similar to the HARPS and NEID data reduction pipelines \citep[see also][]{Bourrier2018}. First, we built a low-resolution version of each telluric-corrected, blaze-preserved EXPRES spectrum by integrating the flux across the central third of each \'echelle order. The wavelength grid for these low-resolution spectra is the mean wavelength in each order. We compared the observed spectrum to the expected spectral energy distribution for a star of similar effective temperature. This nominal spectrum is produced by taking a low-resolution spectrum of a rapidly rotating B-star (constructed in the same manner as for the science spectra), which is taken as a measurement of the total system throughput in ideal conditions. We then multiplied the throughput with a blackbody of the expected \teff{} from the literature. Dividing the reference spectrum by the science spectrum, a sixth-order polynomial was fit to the differences and used to scale the original observed flux.

To obtain the radial velocities, we cross-correlated against an ESPRESSO CCF mask matched to each target according to its spectral type and fit an inverted Gaussian to the resultant CCF. We fit the CCF in a window 1.5 times the full-width at half maximum centered on the CCF minimum, taking the center of the Gaussian as the radial velocity.

\subsection{Archival Radial Velocities}

Archival, out-of-transit radial velocities were compiled for all four systems for inclusion in our joint RM modeling (\S \ref{sec:RM-classic}). These include radial velocities from the Tillinghast Reflector Echelle Spectrograph \citep[TRES,][]{tres_cite}, the FIber-fed Echelle Spectrograph \citep[FIES,][]{fies_cite}, the Extremely High Precision Extrasolar Planet Tracker \citep[EXPERT,][]{expert_cite}, the Tautenburg coude echelle spectrograph \citep[TCES,][]{tces_cite}, and the PRL optical fiber-fed high-resolution cross-dispersed echelle spectrograph \citep[PARAS][]{paras_cite}. Descriptions of the observations and subsequent data reduction can be found in the individual discovery works \citep[][and references therein]{Beatty2012,Pepper2013,Rodriguez2021,Khandelwal2022}. We also provide a summary in Table \ref{tab:archival_rvs}.

\subsection{TESS Photometry}

Photometry from TESS was obtained for all four systems. In all cases, we use the 120 second cadence Presearch Data Conditioning (PDC) flux time series from the Mikulski Archive for Space Telescopes (MAST). The PDC flux is processed by the TESS Science Processing Operations Center (SPOC) pipeline \citep{Jenkins2016}. SPOC-processed photometry was available for KELT-2A in four sectors (43, 44, 45, 71), KELT-3A in two sectors (21, 48), TOI-1333A in two sectors (56, 76), and TOI-1789A in four sectors (44, 45, 46, 48). 

To prepare the flux time series for analysis, we detrended the data with \texttt{W{\={o}}tan} \citep{Hippke} using the sum of cosines method. The window size was tuned for each system to be $\gtrsim 2.2~T_{\text{14}}$ where $T_{\text{14}}$ is the total transit duration \citep{Hippke}. Each sector was detrended individually with a transit mask applied to minimize impact on the transit depth. Only sections of the TESS light curve containing the transit event with one transit duration of baseline on each side were used in the joint modeling.

\subsection{Ground-based Photometry}

\begin{figure}[t!]
    \centering
    \includegraphics[width=0.99\linewidth]{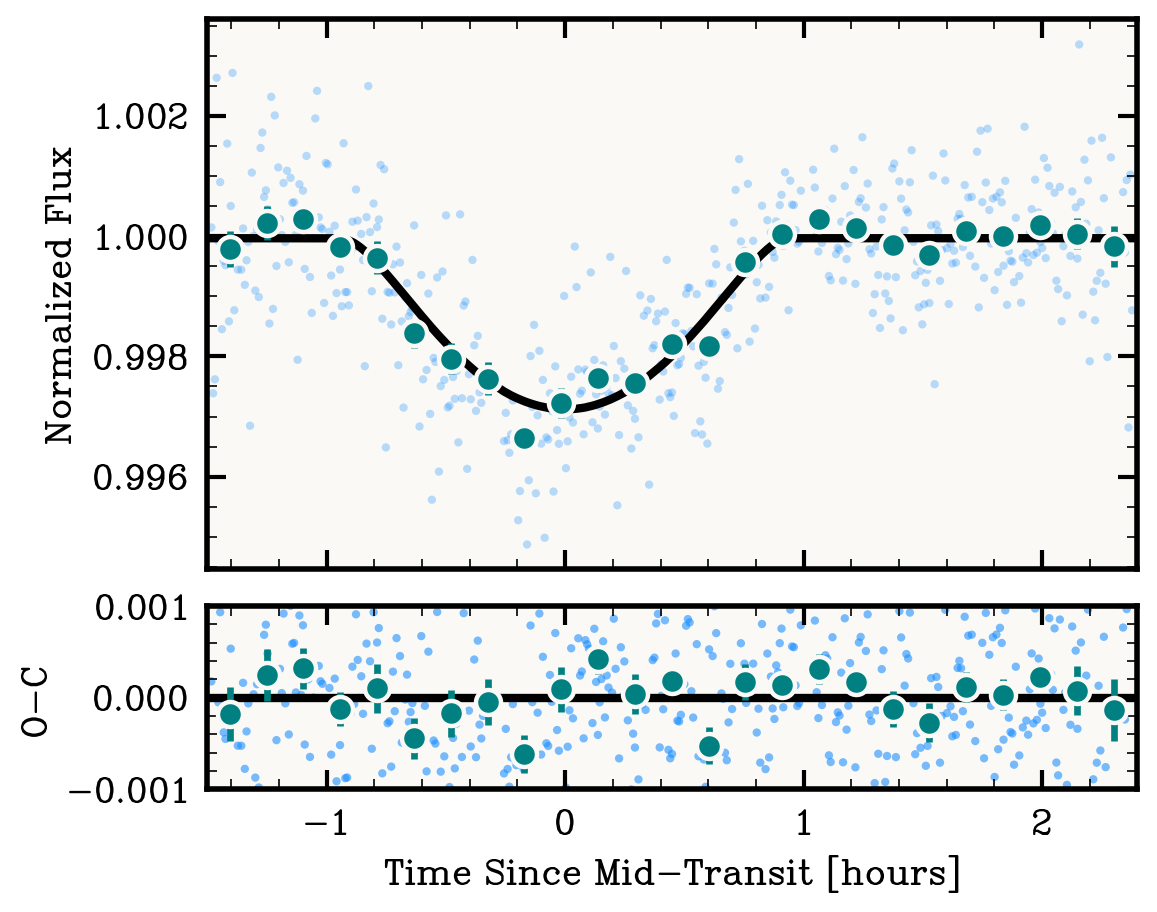}
    \caption{Time-series photometry of TOI-1789 Ab from the Peggy and Eric Johnson 1-meter Telescope. The top panel shows the detrended, normalized photometry with the binned data shown as the larger markers with errorbars. The best fit model is given as the black line. Residuals are given in the bottom panel.}
    \label{fig:pj1m}
\end{figure}

Due to the low expected RM amplitude for TOI-1789 b, we acquired simultaneous photometry from the Peggy \& Eric Johnson Telescope \citep[PJ1M,][]{Hardesty2024} to reduce uncertainty in the mid-transit time ($T_{0}$). PJ1M is a 1-meter corrected Dall-Kirkham from PlaneWave Instruments located at Anderson Mesa Station southeast of Flagstaff, AZ. A Moravian C3-61000 Pro camera mounted on the Nasmyth port provides a 20\farcm63 $\times$ 13\farcm76 field of view. A total of 500 observations were made during the first transit event (2026-03-04 UTC) in the Bessel R-band. An exposure time of 30 seconds was chosen to provide good SNR on both the target and comparison stars. After application of standard flat, bias, and dark corrections, aperture photometry was performed using AstroImageJ \citep{astroimagej}. The reduced lightcurve is shown in Figure \ref{fig:pj1m}.

\section{Analysis}\label{sec:analysis}

\subsection{Modeling the Rossiter-McLaughlin Effect}\label{sec:RM-classic}

For each of the four systems presented here, we modeled the RM effect jointly with TESS photometry, archival radial velocities, and (in the case of TOI-1789 A) ground-based transit photometry. The joint model was produced using the \texttt{exoplanet} package \citep{Foreman-Mackey2021} in combination with the RM model from \cite{Hirano2011} which includes an intrinsic line-broadening parameter, $\beta$, and rotational broadening parameter $\sigma$. We set $\beta$ to the width of the EXPRES resolution element with a normal prior $\mathcal{N}(2.2,0.5)$ \kms. The width of the prior is intended to account for macrotuburlence or other processes that may broaden the line profile. Following \cite{Hirano2010,Hirano2011}, we define $\sigma=$\vsini/1.31. Limb darkening coefficients for each system were calculated using ExoCTK \citep{matthew_bourque_2021_4556063} assuming a tophat bandpass between 0.45 to 0.75 micron and were held fixed. The free parameters used for the transit and radial velocity curves were the orbital period ($P_{\text{orb}}$, time of transit ($T_0$), impact parameter ($b$), scaled planet radius (\rprs), the radial velocity semi-amplitude ($K$), and quadratic limb darkening parameters for TESS and PJ1M. A white noise term, $\sigma$, was also included for each instrument.

\begin{figure*}[t!]
    \centering
    \includegraphics[width=0.9\linewidth]{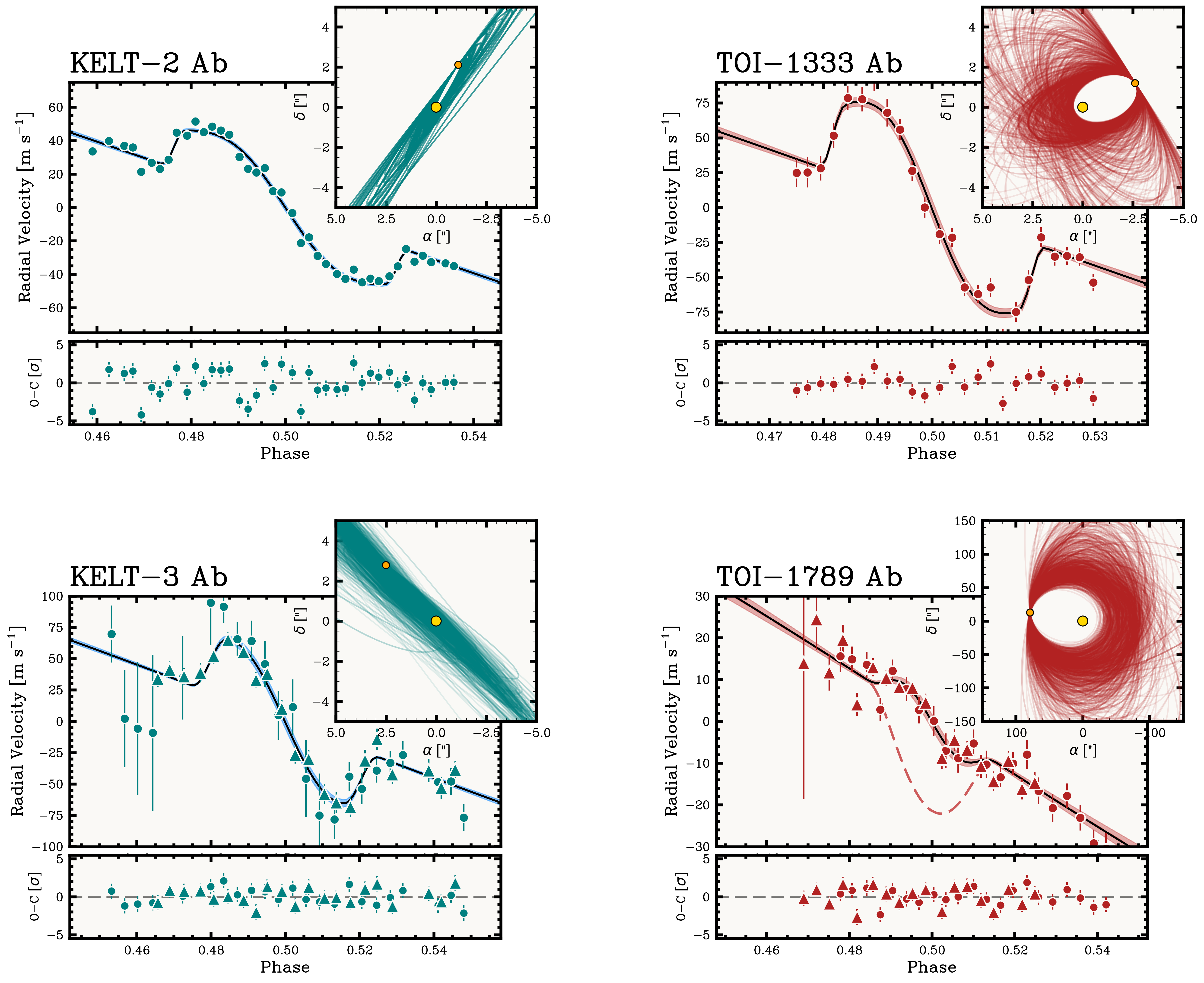}
    \caption{Gallery of the Rossiter-McLaughlin observations presented in this work. In each panel, we show the in-transit radial velocities taken with EXPRES, the median model (black line), and the $\pm$1$\sigma$ confidence intervals (shaded areas). The residuals are given in units of per-datum uncertainty. The insets show the accepted orbits from Bayesian rejection sampling using \texttt{lofti\_gaia} in right ascension ($\alpha$) and declination ($\delta$), as well as the primary (yellow circles) and secondary stars (orange circles). For TOI-1789 Ab, the dashed curve shows the expected Rossiter-McLaughlin curve for a polar orbit. The grazing nature of the transit produces a smaller RM amplitude in an aligned orientation. }
    \label{fig:rm_gallery}
\end{figure*}

The best fit models to the RM data are given in Figure \ref{fig:rm_gallery} and posterior values given in Table \ref{tab:params}. We find all four system to have stellar obliquities consistent with alignment. Our measurements for KELT-3 Ab and TOI-1333 Ab in particular are consistent with the previously reported values in \cite{Knudstrup2024} and \cite{Knudstrup2026}.

%We implement two different parameterizations for the RM model depending on whether the rotational period of the star can be determined, which is true for both TOI-1333A and TOI-1789A and allows for the determination of the true obliquity, $\psi$. In these cases, instead of sampling \vsini~directly, we sample the stellar radius and rotation period to obtain the equatorial velocity $v_{eq}$ \citep{Masuda2020,Stefansson2022}. We then sample the cosine of the stellar inclination to estimate \vsini~through $v_{eq}\sqrt{1-\cos^2{i_{*}}}$,~which accounts for the fact that \vsini~should always be smaller than $v_{eq}$. Finally, the true obliquity is obtained \`a la Equation 9 from \cite{Fabrycky2009}. For KELT-2A and KELT-3A, we simply sample directly in \vsini~and provide the typical projected obliquity.

\subsection{Binary Orbits}

The binary nature of each system considered here has been discussed in previous works and identified through multiplicity searches with \textit{Gaia} \citep{ElBadry2021,PolanskiClark2025}. We characterize the orbits of the stellar companions by first calculating the angle between the sky-projected position and velocity vectors, $\gamma$, given as:

\begin{align}
    \cos{\gamma} = \frac{\vec{r} \cdot \vec{v}}{|\vec{r}||\vec{v}|}
\end{align}

\noindent where $\vec{r}$ and $\vec{v}$ are calculated from positional and proper motion RA and Dec differences, respectively (see \citealt{Tokovinin2015} and \citealt{Rice2024} for further details). The values of $\gamma$ and their associated uncertainties were determined via Monte Carlo sampling following \cite{Rice2024} and fitting a Gaussian to the resulting distribution. 

In addition to $\gamma$, we derive the inclination of the companions' orbit, $i_B$, through 3D orbit fitting using \texttt{lofti\_gaia} \citep{lofti_gaia}, which uses the Bayesian rejection sampling method presented in \cite{Blunt2017} to determine allowed orbits. Masses for the primary stars were taken from the NASA Exoplanet Archive \citep{nea}. Masses for the secondary stars were determined using \texttt{isoclassify} \citep{Huber2017,Berger2023} in grid mode to estimate the the mass from the available \textit{Gaia} parameters. Each \texttt{lofti\_gaia} model fit was run for up to 1,000 accepted orbits. The masses, inclinations, and $\gamma$ values for each companion are given in Table \ref{tab:companion_params}.

\begin{deluxetable}{lcccr}
\label{tab:companion_params}
\tablecaption{Companion characteristics}
\tablehead{\colhead{Companion} & \colhead{Mass [$M_{\odot}$]} & \colhead{$i_B$ [$^{\circ}$]} & \colhead{$\gamma$ [$^{\circ}$] } & \colhead{$s$ [au] }}
\startdata
KELT-2 B & 0.78$\pm0.04$ & 84.2$^{+3.1}_{-2.6}$ & 172.9$\pm0.9$ & 320\\
KELT-3 B & 0.72$\pm0.03$ & 85.5$^{+5.8}_{-6.3}$ & 2.4 $\pm5.6$ & 782 \\
TOI-1789 B & 1.26$\pm0.15$ & 142$\pm12$ & 86.6 $\pm5.5$ & 17845 \\
TOI-1333 B & 0.83$\pm0.07$ & 124$^{+11}_{-12}$ & 73.7$\pm1.1$ & 562
\enddata
\tablecomments{The \textit{Gaia} DR3 identifiers for each component are:\\KELT-2 B: 3438059442854472832\\KELT-3 B: 806492023788937216\\TOI-1789 B: 646124645103549312\\TOI-1333 B: 1978027912379523712}
\end{deluxetable}

\renewcommand{\arraystretch}{1.2}
\begin{deluxetable*}{lcccc}
\label{tab:params}
\tabletypesize{\footnotesize}
\tablecaption{Posterior Values}
\tablehead{\colhead{~~~Parameter Name [units]} &
\colhead{KELT-2 Ab} & \colhead{KELT-3 Ab} & \colhead{TOI-1333 Ab} & \colhead{TOI-1789 Ab}
}
\startdata
\sidehead{\textbf{Transit/RV Parameters}}
P$_{orb}$, Orbital Period [days] & $4.11377605^{+0.00000050}_{-0.00000051}$ & $2.70338926^{+0.00000034}_{-0.00000036}$ & $4.7201872\pm{0.0000028}$ & $3.2087149^{+0.0000029}_{-0.0000028}$ \\
T$_{0}$-2457000.0, Transit Midpoint [days] & $2467.198976^{+0.000053}_{-0.000056}$ & $3732.78593^{+0.00020}_{-0.00021}$ & $3895.85007^{+0.00052}_{-0.00051}$ & $4103.7112\pm{0.0013}$ \\
$b$, Impact Parameter [-] & $0.289^{+0.039}_{-0.046}$ & $0.682^{+0.010}_{-0.011}$ & $0.505^{+0.035}_{-0.041}$ & $0.9856^{+0.0092}_{-0.0098}$ \\
\rprs, Scaled Radius [-] & $0.06825^{+0.00021}_{-0.00022}$ & $0.0952^{+0.00063}_{-0.00078}$ & $0.07526^{+0.00048}_{-0.00044}$ & $0.0743^{+0.0073}_{-0.0070}$ \\
K, RV Semi-ampltidue [\ms] & $158.3^{+5.1}_{-5.2}$ & $181.9^{+5.8}_{-5.3}$ & $223\pm{14}$ & $101.3\pm{5.4}$ \\
$u_{0}$, Limb Darkening (TESS) [-] & $0.196\pm{0.023}$ & $0.153^{+0.056}_{-0.035}$ & $0.214^{+0.062}_{-0.060}$ & $0.24^{+0.13}_{-0.10}$ \\
$u_{1}$, Limb Darkening (TESS) [-] & $0.321^{+0.054}_{-0.049}$ & $0.52^{+0.27}_{-0.28}$ & $0.127^{+0.158}_{-0.090}$ & $0.43^{+0.35}_{-0.30}$ \\
$u_{0}$, Limb Darkening (PJ1M) [-] & - & - & - & $0.117^{+0.131}_{-0.080}$ \\
$u_{1}$, Limb Darkening (PJ1M) [-] & - & - & - & $0.47^{+0.35}_{-0.33}$ \\
\sidehead{\textbf{Rossiter-McLaughlin Parameters}}
$\lambda$, Projected Obliquity [deg] & $0.1\pm{5.0}$ & $0.5^{+2.7}_{-2.6}$ & $0.6^{+4.0}_{-4.2}$ & $0.2\pm{2.9}$ \\
\vsini~, Rotational Velocity [\kms] & $6.58^{+0.33}_{-0.32}$ & $6.96^{+0.44}_{-0.43}$ & $11.2\pm{0.66}$ & $6.81^{+0.47}_{-0.45}$ \\
$\beta$, Line Broadening Parameter [\kms] & $2.25\pm{0.47}$ & $2.25^{+0.49}_{-0.48}$ & $2.32^{+0.48}_{-0.51}$ & $2.33\pm{0.50}$ \\
\sidehead{\textbf{Derived Parameters}}
$a/R_*$, Scaled Orbital Separation [-] & $6.413^{+0.079}_{-0.077}$ & $5.632^{+0.061}_{-0.058}$ & $7.5^{+0.18}_{-0.17}$ & $4.853^{+0.104}_{-0.096}$ \\
$i$, Inclination [$^{\circ}$] & $87.42^{+0.44}_{-0.38}$ & $83.04^{+0.18}_{-0.17}$ & $86.14^{+0.40}_{-0.36}$ & $78.28^{+0.35}_{-0.32}$ \\
\enddata
%\tablenotetext{a}{$\mathcal{N}$ is a normal prior with $\mathcal{N}$(mean, standard deviation)}
%\tablenotetext{b}{$\mathcal{U}$ is a uniform prior with $\mathcal{U}$(lower,upper)}
\end{deluxetable*}

\section{Discussion}\label{sec:discussion}

In the following sections, we compare the spin-orbit alignments with respect to the binary inclinations for the four systems considered in this work and discuss these measurements in the context of the Kraft break. We also examine the obliquities of hot Jupiters in multi-star systems in an attempt to determine the relative contributions of different formation mechanisms to the resulting distribution.

\subsection{Two Fully Aligned Systems}

The angle between the sky-projected position and velocity vectors, $\gamma$, suggests that KELT-2 and KELT-3 are edge-on binaries. Given the low obliquity for both planets, we find KELT-2 and KELT-3 to be \textit{fully aligned}, exhibiting both spin-orbit alignment between the planet and host and orbit-orbit alignment of the stellar components. \cite{Rice2024} reported a relative overabundance of these fully aligned systems, providing evidence for efficient viscous dissipation during the protoplanetary disk phase, which would drive alignment between the star, disk, and binary companion \citep{Zanazzi_VD, Gerbig2024}. In the majority of cases, these were planet hosts with effective temperatures $< 6100$ K. This, along with the tendency for binary inclinations to peak closer to edge-on geometries for cool stars, led \cite{Rice2024} to posit that disks around cooler stars are longer-lived and/or able to dissipate energy more efficiently than disks around hotter stars.

KELT-2 A and KELT-3 A have effective temperatures of 6148$\pm48$ K and 6306$\pm50$ K, respectively, which makes it unlikely that viscous dissipation impacted their formation despite their close projected separations (320 and 782 au, respectively). However, as noted in \cite{Zanazzi_VD}, disk heating likely is not strongly tied to protostar mass \citep{Hayashi1961}, so fully aligned systems may not have a direct dependence on \teff. Instead, the metallicity of the disk may have affected the disk lifetime and heating, as well as the angular momentum of the pre-main sequence star. For example, \cite{Bitsch2015} presented a disk model with careful treatment of heating and cooling effects and found that disks with higher metallicity are hotter. On the other hand, \cite{Gehrig2023} showed that stars with higher stellar luminosities at lower metallicities have hotter disks. This contributes to low-metallicity systems having shorter disk lifetimes and faster rotation during the pre-main sequence phase, potentially leading to less efficient viscous dissipation. In either case, disk metallicity affects the effectiveness of viscous dissipation; it is most efficient for cooler disks and larger external torques \citep{Zanazzi_VD}.

Both KELT-2 A and KELT-3 A have near-Solar metallicities: 0.03$\pm$0.07 and 0.04$\pm$0.08, respectively \citep{Bonomo2017}. The fully aligned hot Jupiters considered in \cite{Rice2024} are also all near Solar metallicity, below the peak of the hot Jupiter metallicity distribution \citep[0.10$\pm$0.012,][]{Osborn2020}, with some exceptions. The metallicity of HAT-P-1 A is super-Solar, but with a projected binary separation of 1,800 au, viscous dissipation may have been less effective. HAT-P-22 A also shows super-Solar metallicity. However, this system has a mass ratio of $m_{\text{p}}/M_{\star} = 2.2\times10^{-3}$, placing it in a regime where close-in planets are aligned regardless of host star \teff~\citep{Rusznak2025}. TrES-4 A stands out as the highest-metallicity hot Jupiter host in a fully aligned system, with [Fe/H] = 0.28$\pm$0.09 \citep{Bonomo2017}. 

The remaining two systems in our sample, TOI-1333 Ab and TOI-1789 Ab, have near face-on binary orbits, in contrast to KELT-2 Ab and KELT-3 Ab. These perpendicular geometries are consistent with previous results from \cite{Behmard2022}, which showed that binaries hosting close-in giants tend to be face-on, though \cite{Rice2024} disputes this preference. The small projected separation of 560 au places TOI-1333 Ab within the regime of viscous dissipation, so it is surprising that it is not fully aligned. The \teff~of the host is above the 6100 K boundary, but its super-Solar metallicity points to disk composition having potentially influenced the observed geometry. TOI-1789 has the largest separation in our sample ($\sim18,000$ au), suggesting it is likely outside the regime of the viscous dissipation mechanism. TOI-1789 Ab likely migrated and tidally circularized its orbit in a way typical for hot Jupiters around cool, single stars \citep{Rice2022,Winn2010b}. 

\subsection{Revision of the Multi-Star Realignment Break}\label{sec:kraft_break}

The Kraft break is a transition in stellar structure between stars with outer convective envelopes with low rotational velocities and those that are rapid rotators with mostly radiative envelopes at $\sim6500$ K \citep{Kraft1967}. This transition is seen also in the obliquity distribution of hot Jupiters, with planets orbiting cooler stars showing predominantly low obliquities, while those around hot stars show a wider distribution of spin-orbit misalignment \citep{Winn2010b,Albrecht2022}. In the following section, we reserve the term ``Kraft break'' to refer to the transition seen in \vsini~measurements, and introduce the term ``realignment break" (\trb) to refer to the \teff~where the obliquity distribution of hot Jupiters changes.

Recently \cite{Wang2026} identified a different realignment break for hot Jupiters in multi-star systems. They find that the transition from aligned to misaligned occurs at $6105_{-133}^{+123}$ K, whereas for single stars, the boundary is closer to the Kraft break derived from \vsini~measurements \citep[see][]{Beyer2024}. Given that the \teff~values of the planet hosts in our sample straddle this boundary, we follow a similar methodology of using a Kolmogorov-Smirnov (KS) scan to revise the binary realignment break estimate using the obliquities presented here\footnote[1]{A previous obliquity measurement for KELT-3 Ab was included in the realignment break determination from \cite{Wang2026}. For our determination of \trb, we use our obliquity measurement.}, as well as the recent measurement made for HAT-P-57 Ab \citep{Balkoova2026}. In \cite{Wang2026}, the bootstrapped Kolmogorov-Smirnov scan resulted in a distribution of \teff~cuts dominated by a peak near 6100 K, though the distribution also included smaller peaks on either side of this peak. The cooler peak is driven by the highly misaligned KELT-23 Ab \citep{Giacalone2025}, and the hotter peak results from including host stars with \teff~$>$ 6600 K; these were excluded in \cite{Wang2026}. Making similar cuts, we obtain a value of \trb$=6128^{+115}_{-128}$ K, which is slightly higher than, but consistent with, the value from \cite{Wang2026}.

We also experiment with a stricter sample. Specifically, we remove two stars, WASP-22 and WASP-72, that were included based on a Renormalized Unit Weight Error (RUWE) value greater than 1.4, which can indicate the presence of a companion, but that have not had a companion directly identified. Two more stars, HAT-P-50 and WASP-103, have both a high RUWE value and potential companion detections from speckle imaging, but the boundedness of the companions has not been confirmed. Additionally, K2-237 is excluded, as its co-proper motion index (CPM) of 3.4 is both close to the threshold of CPM$>$3 used in \cite{EelesNolle2025} and far below the median CPM of 86 for the other \textit{Gaia}-resolved multiples in our sample. Removal of these stars (along with the cuts mentioned previously) yields a Kraft break value of \trb$=6156^{+103}_{-130}$ K.

\begin{figure*}[t!]
    \centering
    \includegraphics[width=0.98\linewidth]{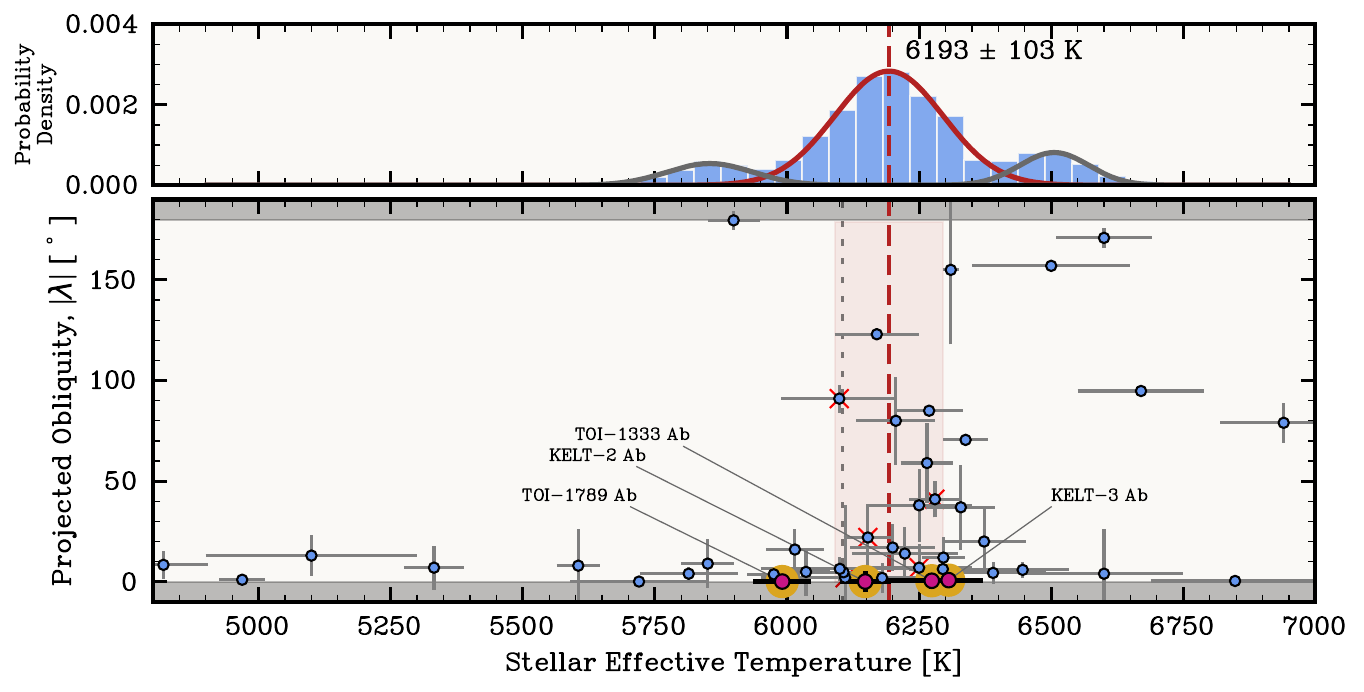}
    \caption{Stellar obliquities of hot Jupiters in multi-star systems from \cite{Wang2026}. The obliquity measurements presented in this work are shown as the larger, yellow points. A red `X' indicates measurements that were left out of the final derivation of the \teff~boundary due to either not having a companion directly identified or the co-proper motion not being confirmed. The red dashed line and shaded area show the mean and standard deviation of the realignment break (\trb) determined through the bootstrapped Kolmorgorov-Smirnov scan shown in the top panel. The gray dashed line is the multi-star realignment break derived by \cite{Wang2026}.}
    \label{fig:kraft_break}
\end{figure*}

Including stars with \teff~$>$ 6600 K and KELT-23 Ab results in two smaller peaks near 5850 K and 6500 K (Figure \ref{fig:kraft_break}). Accepting these as features of the sample under consideration, we fit the distribution with a three-component Gaussian model, which returns a ``primary'' \trb~value at $6193\pm103$ K.

The overall result is that the multi-star Kraft breaks shows a degree of dependence on how the sample is curated and, while we find hotter \trb{} values, they are within the confidence interval of the original work. Since the 6193 K value is derived with a more complete dataset, we adopt this value as the boundary between hot and cool hosts for multi-star systems for the remainder of the work.

\subsection{Formation Mechanism Contributions for Hot Jupiters in Multi-Star Systems}

The population of close-in giants we observe today is likely sculpted by a variety of formation mechanisms \citep{Fabrycky2009,Naoz2012,Rice2022,Dong2023}. The distribution of obliquities for hot Jupiter systems implies that many of these planets experienced dynamically hot migration histories, such as being launched into an excited orbit and experiencing subsequent high-eccentricity migration \citep{Fabrycky2007,Nagasawa2008,Petrovich2015}. On the other hand, less chaotic mechanisms \citep[i.e. disk migration,][]{Goldreich1980} may also play a role, as evidenced by some hot Jupiters having interior small planets \citep{Hord2022,Korth2023,Korth2024,Grieves2025,Quinn2026}. Previous works have attempted to quantify the relative contribution of different migration methods by comparing the expected $\lambda$ distribution with what is observed.

For planets in binary systems, the von-Zeipel-Kozai-Lidov (ZKL) mechanism may be naturally invoked to explain the observed misalignments. The distribution of true spin-orbit angles ($\psi$) resulting from ZKL migration has been studied extensively and suggests a wide distribution of $\psi$ with peaks near $\sim35^{\circ}$ and $\sim110^{\circ}$ \citep{Fabrycky2007,Anderson2016}. When examining the population of hot Jupiters with hosts above the Kraft break derived in \S \ref{sec:kraft_break}, we find that aligned systems\footnote[2]{Defined as systems with $|\lambda|<10^{\circ}$ to at least 1$\sigma_{\lambda}$} are more prevalent above the \teff~break for multi-star systems than in single-star systems (20\% vs. 0\%, respectively), suggesting additional mechanisms are at work.

To compare the observed obliquity distribution to what we expect for a given migration mechanism (such as ZKL), we implement a simple population synthesis model to compare the observed obliquities to a population with an assumed underlying distribution. We first sample a true spin-orbit angle, $\psi$, then randomly select an argument of periapse from a uniform distribution [0, $2\pi$), a stellar inclination from a uniform distribution $\cos{i_{\star}}$, and an orbital separation $a/R_{\star}$ from a uniform distribution [2.5, 10). These values are used to calculate the orbital inclination and impact parameter of the planet. If the impact parameter is less than 0.9, the planet is considered transiting and the projected obliquity is recorded. Systems are simulated until we obtain a set of 23 transiting planets matching the number of observed systems with an obliquity measurement and a host \teff~$>6193$ K. This was repeated for 10,000 iterations. By considering systems only around hot stars in this simplified framework, we assume that the tidal realignment is inefficient and that the distribution of observed obliquities is reflective of the post-migration alignment distribution. 

\begin{figure}[t!]
    \centering
    \includegraphics[width=0.98\linewidth]{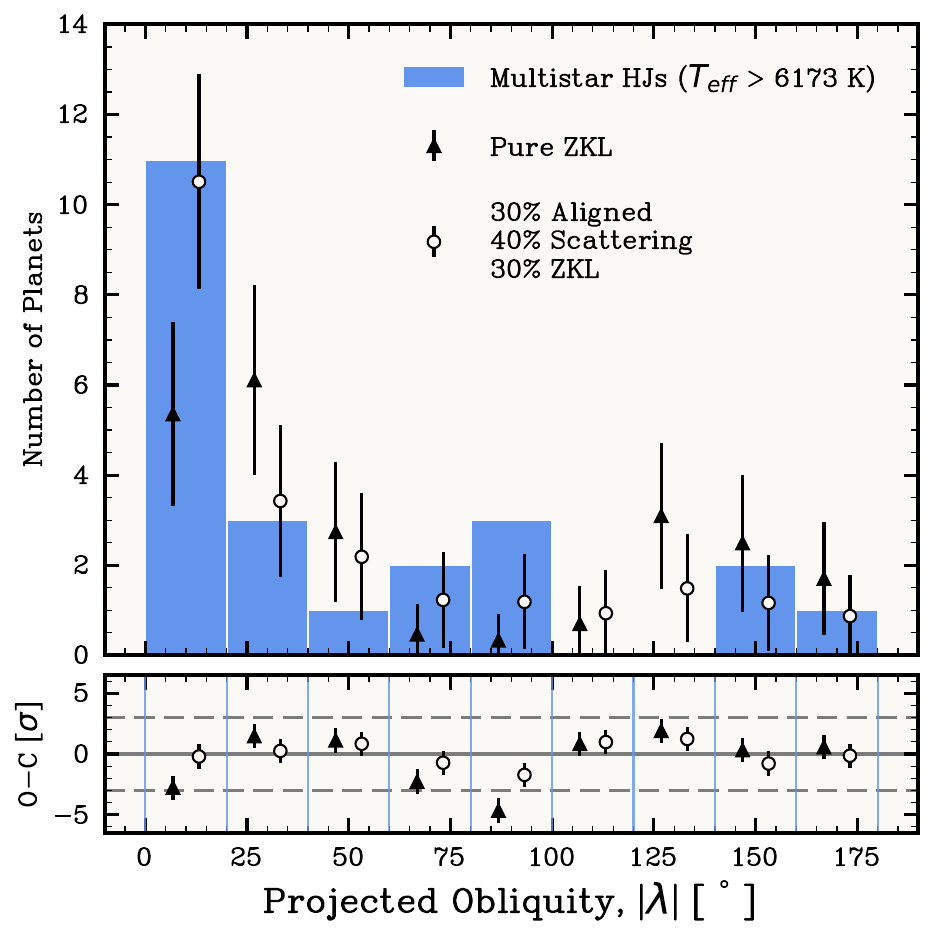}
    \caption{Stellar obliquities of hot Jupiters in multi-star systems with host star \teff~$>$6193 K. The error bars are derived from various population synthesis simulations for a von-Zeipel-Kozai-Lidov-driven population (triangles) and for a mixture model of different source populations (open circles). The bottom panel shows the residuals with respect to the histogram heights in units of standard deviation.}
    \label{fig:pop_synth_hist}
\end{figure}

We first test the case of a population formed through ZKL migration alone. For the underlying $\psi$ distribution we use the distribution from \cite{Anderson2016} most appropriate to the sample under consideration: namely, one that simulates a population of hot Jupiters around F-type stars under the influence of a stellar companion at various separations. The results are given in Figure \ref{fig:pop_synth_hist}. These simulations are generally a poor fit to the observed obliquity distribution for hot Jupiters around hot stars. In particular, the number of aligned systems is underestimated by $\sim3\sigma$. The frequency of aligned systems also falls rapidly past $\lambda\sim15^{\circ}$. Furthermore, we observe more near-polar systems than would be predicted by an underlying distribution driven by ZKL alone by nearly 5$\sigma$.

In an attempt to better replicate the observed distribution, we run a series of simulations that includes contributions from a fully aligned population, such as one that we may expect from either disk migration or coplanar high eccentricity migration \citep{Petrovich2015}, as well as a population that resembles planet-planet scattering \citep{Nagasawa2008}. We iterate over a grid of relative contribution fractions of step size 0.1. For each instance, we compute the Kolmogorov-Smirnov distance (D) between the simulated and observed distributions. This produces a set of 10,000 D statistics from which we calculate the mean and standard deviation for each combination of contribution fractions.

\begin{figure}[t!]
    \centering
    \includegraphics[width=\linewidth]{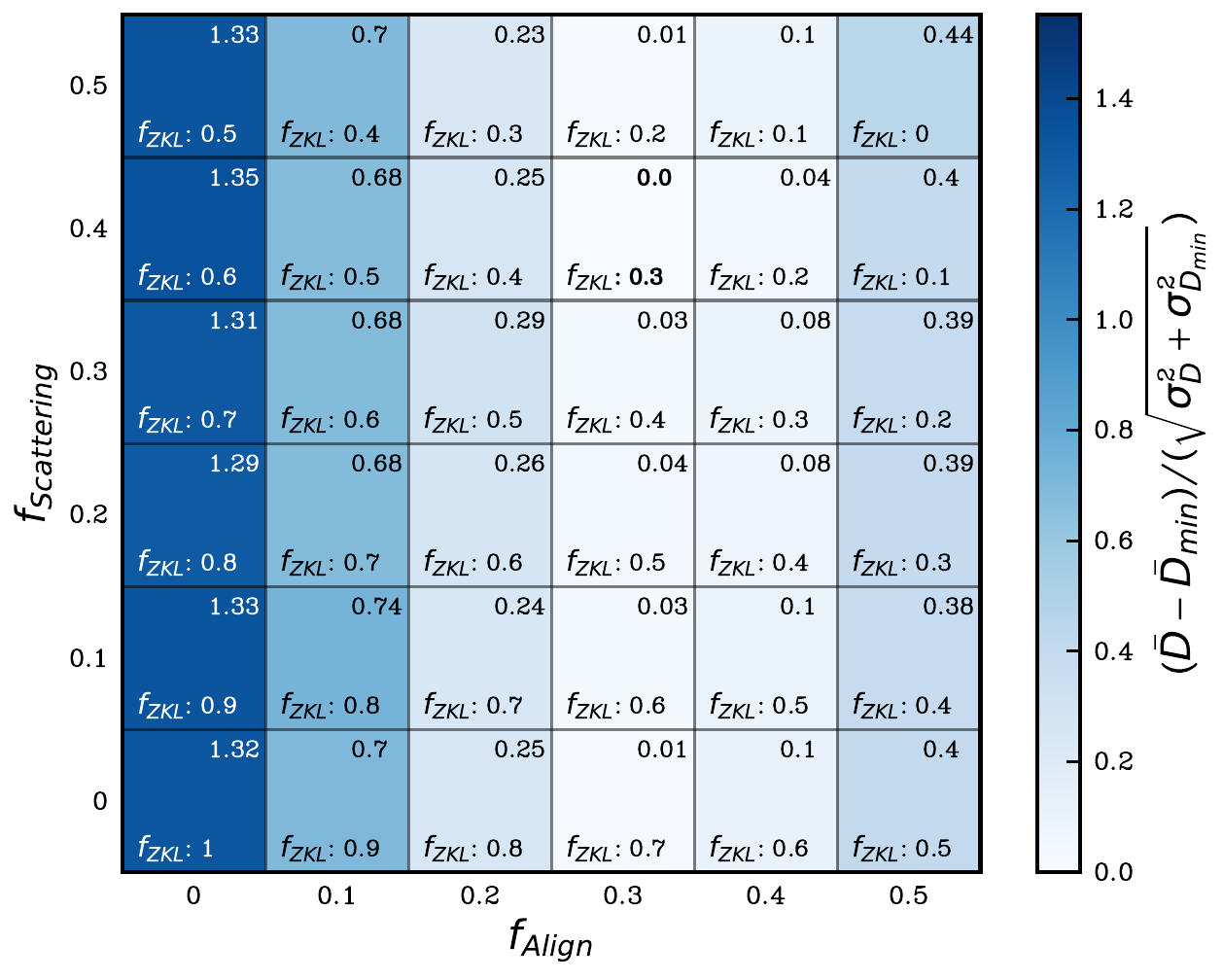}
    \caption{Grid of tested contributions used in our population synthesis simulations. Each tile represents a combination of populations that are aligned ($f_{\text{Align}}$), planet-planet scattered ($f_{\text{Scattering}}$), and driven by the von-Zeipel-Kozai-Lidov mechanism ($f_{\text{ZKL}}$), such that $f_{\text{Align}} + f_{\text{Scattering}} + f_{\text{ZKL}} = 1$. The tiles are color coded according to the difference between the mean Kolmogorov-Smirnov distance of each combination ($\bar{D}$) and the overall minimum of all tested combinations ($\bar{D}_{min}$), normalized by the quadrature sum of the standard deviations. In each tile, this statistic is given in the upper right. }
    \label{fig:pop_synth_grid}
\end{figure}

Figure \ref{fig:pop_synth_grid} shows the grid of tested contribution fractions. Overall, it is difficult to distinguish between the majority of combinations, likely due to small sample size, though it is clear that a significant fraction of aligned systems (between 30-40\%) is needed to reproduce the observed $\lambda$ distribution. The ZKL contribution fraction that matches the current sample best is 30\%, in agreement with previous studies \citep{Naoz2012}, but is degenerate with planet-planet scattering. If scattering events are rare, ZKL may yield up to 70\% of the hot Jupiters in multi-star systems, though some mechanism is needed to produce the excess of polar planets that the scattering contribution helps to replicate. 

The formation rate of hot Jupiters through ZKL has been estimated to be between 0-5\%, with many more planets undergoing migration but becoming tidally disrupted \citep{Anderson2016,Munoz2016}. As such, a pure ZKL population may not match the observed obliquity distribution even in multi-star systems making ZKL less of a ``smoking gun" formation mechanism for hot Jupiters in binaries.

The large fraction of aligned systems is worthy of note in context of the previous section's results. Fundamentally, the transition in the hot Jupiter obliquity distribution reflects a transition in stellar structure, as well as how close-in giant planets tidally interact with their hosts \citep{Kraft1967,Winn2010b}. How a stellar companion influences these tidal interactions is not clear, but the KS scan shows a break at $\sim6,500$ K, which is close to the break seen in single stars. The systems in this sample should have evolved effectively as single stars given the separations of $\sim$thousands of au, and the presence of the secondary peak in Figure \ref{fig:kraft_break}. The $\sim6,200$ K boundary suggests, however, that the companions in a subset of these systems are acting to suppress tidal realignment. The relatively sharp boundary coincides with a slight uptick in the \vsini~distribution for stars with companions given in \cite{Beyer2024} and \cite{Wang2026}, implying that this may also be linked to a difference in stellar structure. 

The treatment of hot Jupiters in multi-star systems as a separate population raises questions about the impact of stellar companions on planet formation. Which properties of protoplanetary disks influence the formation of fully aligned systems? Is ZKL the dominant migration mechanism? Is the obliquity distribution shaped by more than one process? Addressing these questions, especially given the significantly wider parameter space the presence of one or more companions opens up, underscores the need for many more spin-orbit measurements of S-type systems. 

\section{Summary}

In this first CHIANTE work, we presented Rossiter-McLaughlin measurements from EXPRES of four Ariel Tier 3 targets: KELT-2 Ab, KELT-3 Ab, TOI-1333 Ab, and TOI-1789 Ab. All four are S-type systems, and the hot Jupiter orbits the primary star. We performed a joint fit to in-transit EXPRES and archival RVs along with TESS photometry and find that these planets are well-aligned with their host stars. This implies that the planets either underwent a smooth migration to their current positions, or they migrated in a fashion that increased their obliquity and subsequent tidally realignment. KELT-2 Ab and KELT-3 Ab are fully aligned in the plane of the sky, showing both spin-orbit alignment of the planet with the host star and orbit-orbit alignment of the stellar companions. This adds to the population of fully aligned systems identified in \cite{Rice2024}, which may be a result of viscous dissipation during the protoplanetary disk phase. 

The stellar \teff~values of our four targets straddle the multi-star realignment break identified in \cite{Wang2026}; we revise this value to include the measurements presented in this work. We derive a boundary that is higher than, although consistent with, the one found previously. The exact derived location of this transition depends on how the sample is constructed; ultimately we adopt a value of \trb$=6193\pm103$ K. 

We examine the current obliquity distribution of hot-Jupiter-hosting stars in multi-star (including triples and binaries) systems and find an excess of aligned systems above the \teff~boundary. Using a toy population synthesis model, we compare the observed $\lambda$ distribution to what would be expected for various migration mechanisms. We find that a population driven purely by von-Zeipel-Kozai-Lidov oscillations is inconsistent with the current sample of stellar obliquities. A mixture of populations -- one that is preferentially aligned, one from ZKL, and one arising from planet-planet scattering -- replicates the observed distribution better, but only places weak constraints on the contribution of ZKL relative to planet-planet scattering.

\section{Data Availability}

TESS data is publicly available on the Mikulski Archive for Space Telescopes (MAST). Timeseries photometry from PJ1M and EXPRES radial velocities are archived on the ExoFOP-TESS webpage for each respective star. EXPRES spectra and PJ1M full-frame images are available upon request to the corresponding author.

\begin{acknowledgements}

A.S.P thanks Diego Mu\~noz, Taylor Kutra, Claudia Danti, Kathryn Turrentine and Christine Reid for their insightful conversations that greatly improved the quality of this manuscript.

A.S.P acknowledges support through the Percival Lowell Postdoctoral Fellowship which is funded, in part, through a generous donation from John and Ginger Giovale. He also thanks the Lowell TAC for the time needed to conduct these observations. A.S.P thanks Atlas for the accompaniment during the nights of observations for which this data was taken. This work was supported by NASA grant 24-USCAPS24-0012.

M.R. acknowledges support from Heising-Simons Foundation Grants \#2023-4478 and \#2023-4655, as well as NASA grant \#80NSSC24K0359. 

These results made use of the Lowell Discovery Telescope (LDT) at Lowell Observatory.  Lowell is a private, non-profit institution dedicated to astrophysical research and public appreciation of astronomy and operates the LDT in partnership with Boston University, the University of Maryland, the University of Toledo, Northern Arizona University and Yale University. Lowell Observatory sits at the base of mountains sacred to tribes throughout the southwest United States. We honor their past, present, and future generations, who have lived here for millennia and will forever call this place home.

The EXPRES team acknowledges support for the design and construction of EXPRES from NSF MRI-1429365, NSF ATI-1509436 and Yale University. DAF gratefully acknowledges support to carry out this research from NSF 2009528, NSF 1616086, NSF AST-2009528, the Heising-Simons Foundation, and an anonymous donor in the Yale alumni community.

\end{acknowledgements}

\facilities{TESS, LDT (EXPRES), PJ1M}

\appendix

\section{Radial Velocities}

\renewcommand{\arraystretch}{1.2}
\begin{deluxetable}{lccc}[h]
\label{tab:rv_sample}
\tabletypesize{\footnotesize}
\tablecaption{EXPRES Radial Velocities}
\tablehead{
\colhead{Target} 
& \colhead{Time [TBJD]}
& \colhead{RV [\ms]} 
& \colhead{$\sigma_{\text{RV}}$ [\ms]}
}
\startdata
KELT-2 A & 3679.82838 & -47374.64 & 6.26\\
KELT-2 A & 3680.61046 & -47467.09 & 8.55\\
KELT-2 A & 3680.61838 & -47479.17 & 7.96\\
KELT-2 A & 3680.62661 & -47497.36 & 9.31\\
KELT-2 A & 3680.63398 & -47503.00 & 8.03\\
... & ... & ... & ... \\
KELT-3 A & 3732.71228 & 27869.71 & 6.49\\
KELT-3 A & 3732.72418 & 27872.77 & 7.81\\
KELT-3 A & 3732.73389 & 27886.06 & 6.35\\
KELT-3 A & 3732.74421 & 27899.00 & 5.78\\
KELT-3 A & 3732.75550 & 27889.41 & 5.77\\
... & ... & ... & ... \\
TOI-1333 A & 3895.73238 & -15613.27 & 10.06\\
TOI-1333 A & 3895.74176 & -15612.93 & 10.91\\
TOI-1333 A & 3895.75340 & -15609.93 & 8.93\\
TOI-1333 A & 3895.76455 & -15586.44 & 8.93\\
TOI-1333 A & 3895.77685 & -15559.46 & 8.53\\
... & ... & ... & ... \\
TOI-1789 A & 4103.64921 & -41309.65 & 3.06\\
TOI-1789 A & 4103.66064 & -41310.89 & 3.11\\
TOI-1789 A & 4103.67108 & -41321.66 & 2.75\\
TOI-1789 A & 4103.68064 & -41312.41 & 2.71\\
TOI-1789 A & 4103.69144 & -41316.75 & 2.91\\
... & ... & ... & ... \\
\enddata
\tablecomments{This table is available in its entirety in a machine-readable form.}
\tablecomments{TBJD denotes TESS barycentric Julian date, i.e. BJD-2457000.0.}
\end{deluxetable}

\begin{deluxetable*}{lcccc}[t!]
\label{tab:archival_rvs}
\tablecaption{Summary of Archival Radial Velocities}
\tablehead{\colhead{Target} &
\colhead{Instrument} & \colhead{No. Spectra} & \colhead{Date Range [UTC]} & \colhead{RV Precision [\ms]}
}
\startdata
KELT-2 A & TRES & 17 & 2012-02-01 to 2012-05-04 & 31$\pm$15\\
\hline
%%%
\multirow{3}{*}{KELT-3 A} & TRES & 15 & 2012-02-26 to 2012-05-01 & 16$\pm$4 \\
& FIES & 5 & 2012-03-13 to 2012-03-17 & 20$\pm$4 \\
& EXPERT & 5 & 2012-05-02 to 2013-01-23 & 26$\pm$3 \\
\hline
%%%
TOI-1333 A & TRES & 16 & 2019-10-21 to 2019-11-05 & 40$\pm$7\\
\hline
%%%
\multirow{2}{*}{TOI-1789 A} & PARAS & 16 & 2020-12-20 to 2021-03-19  & 30$\pm$8 \\
& TCES & 21 & 2021-02-22 to 2021-04-05 & 30$\pm$7 \\
\enddata
\end{deluxetable*}

\bibliography{main}{}
\bibliographystyle{aasjournalv7}

\end{document}